# Characterizing the Instrumental Profile of LAMOST

Qian Liu (刘倩)[1,2], Zhongrui Bai (白仲瑞)[1,7], Ming Zhou (周明)[3], Mingkuan Yang (杨明宽)[1,2,4,5], Xiaozhen Yang (杨肖振)[1,2], Ziyue Jiang (蒋子悦)[1,2], Hailong Yuan (袁海龙)[1], Ganyu Li (李甘雨)[1,2], Yuji He (何玉吉)[1,6], Mengxin Wang (汪梦欣)[1], Yiqiao Dong (董义乔)[1], and Haotong Zhang (张昊彤)[1,7]

[1] Key Laboratory of Optical Astronomy, National Astronomical Observatories, Chinese Academy of Sciences, Beijing 100101, People's Republic of China; zrbai@bao.ac.cn, htzhang@bao.ac.cn

[2] School of Astronomy and Space Science, University of Chinese Academy of Sciences, Beijing 100049, People's Republic of China

[3] College of Photonics and Optical Engineering, Aerospace Information Technology University, Jinan 250299, People's Republic of China

[4] International Centre of Supernovae (ICESUN), Yunnan Key Laboratory of Supernova Research, Yunnan Observatories, Chinese Academy of Sciences, Kunming 650216, People's Republic of China

[5] Key Laboratory for Structure and Evolution of Celestial Objects, Chinese Academy of Sciences, Kunming 650216, People's Republic of China

[6] Leibniz-Institut für Astrophysik Potsdam (AIP), An der Sternwarte 16, 14482 Potsdam, Germany

## Abstract

The instrumental profile (IP) of a telescope is of great significance for spectroscopic analyses, especially for wavelength calibration and stellar parameter measurements. The Large Sky Area Multi-Object Fiber Spectroscopic Telescope (LAMOST) employs arc lamps for wavelength calibration. These lamps produce sharp emission lines with known wavelengths, and the observed arc lamp spectra can well characterize the IP. However, IPs are influenced by multiple factors, making them difficult to model accurately with traditional methods. Neural networks, which can automatically capture complex patterns and nonlinear features in data, provide a promising approach for high-precision IP measurement. We therefore construct a multilayer perceptron based on The Payne neural network to derive IPs for LAMOST. After training, the model can retrieve the IP for any fiber, at any wavelength, and at any time. We then apply the derived IP to stellar radial velocity (RV) measurements and analyze the impact of different IP center localization methods on the results. Finally, the dispersion of the measured RVs is reduced by approximately 3 km s$^{-1}$. This improvement will facilitate the search for long-period binary stars via RV variations.

## 1. Introduction

The instrumental profile (IP) of a spectrograph describes how an intrinsically narrow (monochromatic) spectral line is broadened and possibly distorted as it passes through the instrument. In astronomical spectroscopic analysis, the wavelength calibration is often performed by determining the center of the IP (F. Zhao et al. 2019; T. M. Schmidt et al. 2021; T. M. Schmidt & F. Bouchy 2024). To measure various parameters of the stars, convolving template spectra with the IP to match the instrumental resolution is necessary (T.-O. Husser et al. 2013; A. Lançon et al. 2021). Thus, accurate characterization of the IP is crucial as even small deviations in its shape can introduce biases in the derived quantities like radial velocity (RV; T. Hirano et al. 2020) and the fine-structure constant (F. Berlfein et al. 2025).

The shape of the IP is influenced by a variety of factors, including the condition of the instrument itself—such as the optical system and the detector response ( P. Antilogus et al. 2014; A. A. Plazas et al. 2016)—as well as environmental conditions, like temperature and humidity (W. H. De Vries et al. 2007; C. Chang et al. 2012; T. I. Liaudat et al. 2023). Over the past several decades, a variety of methods have been developed to model the IP of spectrographs using calibration sources since they become broadened or distorted after passing through the instrument:

$$I_{\rm obs}(\lambda) = \int_{-\infty}^{+\infty} \delta(\lambda' - \lambda_0) \cdot h(\lambda - \lambda')\, d\lambda'. \quad (1)$$

The observed profile $I_{\rm obs}(\lambda)$ is the convolution of the intrinsic emission line profile, which is a Dirac delta function $\delta(\lambda' - \lambda_0)$, and the IP $h(\lambda - \lambda_0)$. Equation (1) simplifies to

$$I_{\rm obs}(\lambda) = h(\lambda - \lambda_0). \quad (2)$$

Usually, the calibration source's intrinsic linewidth is far narrower than the IP's width, the approximation (i.e., Equation (2)) holds, and the observed spectra provide ideal data for extracting the IP.

There are many examples of extracting the IP using calibration lamp sources. Iodine absorption cells became mature as calibration sources in the early 1990s, and most of the early efforts relied on them, e.g., the Hamilton Echelle Spectrograph at Lick Observatory (G. W. Marcy & R. P. Butler 1992; R. P. Butler et al. 1996), the Subaru High Dispersion Spectrograph, and the Okayama HIDES (E. Kambe et al. 2002). Since the laser frequency comb (LFC) generates a dense, broadband spectrum composed of many narrow lines equally spaced in frequency, it has been increasingly adopted for IP characterization in subsequent studies, e.g., the Subaru InfraRed Doppler spectrograph (T. Hirano et al. 2020), the high-resolution spectrograph at the Xinglong 2.16 m Telescope (F. Zhao et al. 2019; Z. Hao et al. 2020), and the ESPRESSO high-resolution spectrograph at the Very Large Telescope (D. Milaković & P. Jethwa 2024; T. M. Schmidt & F. Bouchy 2024), following its first astronomical application in 2008 (C.-H. Li et al. 2008). Other calibration sources, such as hollow-cathode lamps, have also been used, e.g., the GHRS at the Hubble Space Telescope (J. A. Cardelli et al. 1990).

[7] Corresponding authors.

With regard to the modeling strategies, some studies adopt parametric approaches. They typically assume analytic functional forms, such as combinations of Gaussian and power-law functions (J. A. Cardelli et al. 1990), Gaussian functions with satellite Gaussians (J. A. Valenti et al. 1995), sums of multiple Gaussians (G. W. Marcy & R. P. Butler 1992; R. P. Butler et al. 1996; F. Zhao et al. 2019), or backbone-plus-residual models (Z. Hao et al. 2020). Because the IP is influenced by multiple factors, it is often difficult to determine an appropriate functional form or combination of functions to describe it. Although parametric models can be extended to incorporate multiple variables, their performance depends strongly on the assumed functional form and may become increasingly restrictive when attempting to capture complex, nonlinear interactions among these factors. In contrast, nonparametric approaches relax explicit assumptions about the IP shape and reconstruct it directly from the data. These methods typically represent the IP as numerical vectors or infer it using machine learning techniques such as Gaussian process regression (T. Hirano et al. 2020; D. Milaković & P. Jethwa 2024; T. M. Schmidt & F. Bouchy 2024). Consequently, machine learning–based frameworks provide a flexible, data-driven way to capture high-dimensional and nonlinear dependencies, which is particularly advantageous for modeling IP variations arising from multiple interacting effects. Together, these methods provide valuable insights into the IP characterization of various spectrographs.

Compared to the above classical single-object, high-resolution spectrographs, multifiber systems may present greater challenges for IP characterization: precise and potentially distinct IP models must be established for hundreds or even thousands of fibers simultaneously. For example, a two-dimensional (2D) line-spread function (LSF) model varying with both wavelength and fiber position was developed for the spectrographs of the Sloan Digital Sky Survey (SDSS) and the Baryon Oscillation Spectroscopic Survey (BOSS; S. A. Smee et al. 2013). In that work, arc lamp emission lines were fitted with Gaussian functions to measure their widths, which were then modeled as a function of wavelength using Legendre polynomials. The spatial variation of the LSF was further characterized as a function of fiber position across the detector. Similarly, D. R. Law et al. (2021) constructed wavelength- and fiber-dependent LSF models for the SDSS-IV MaNGA survey, which employs BOSS spectrographs coupled with integral field unit (IFU) fiber bundles (K. Bundy et al. 2015; M. R. Blanton et al. 2017). Their models were derived from arc lamp and night-sky emission lines, with Gaussian fits used to describe individual line profiles and polynomial functions adopted to parameterize their wavelength and fiber dependence. The resulting three-dimensional LSF data cubes were matched to each reconstructed MaNGA galaxy data cube. Their pipeline LSF achieved an overall accuracy at the $\sim$1% level, enabling reliable measurements of velocity dispersions down to $\sim$20 km s$^{-1}$.

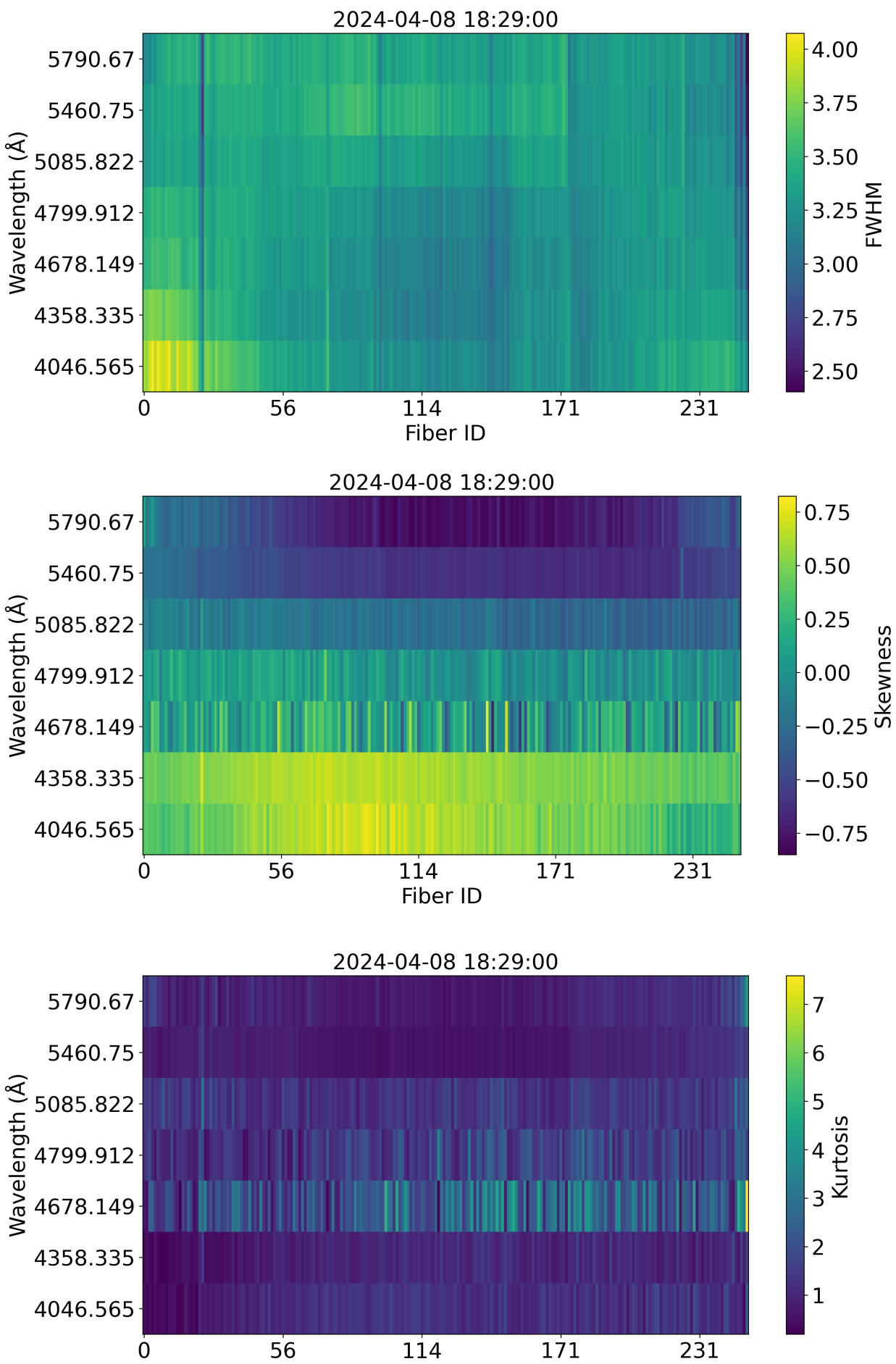

**Figure 1.** This figure shows the variation of the lamp line profile parameters for LAMOST Spectrograph No. 1 at a given time (indicated in the title of each panel). From top to bottom, the panels show 2D maps of the FWHM, skewness, and kurtosis of the profiles, with fiber ID on the $x$-axis and wavelength on the $y$-axis.

For a telescope like the Large Sky Area Multi-Object Fiber Spectroscopic Telescope (LAMOST), accurate IP characterization is also a highly nontrivial task. LAMOST is a large-aperture, wide-field spectroscopic survey telescope located in China. It is equipped with 16 spectrographs, each feeding 250 fibers, enabling simultaneous observations of up to 4000 objects and delivering exceptionally high survey efficiency (X.-Q. Cui et al. 2012). Although LAMOST has already released an enormous volume of spectra, studies dedicated to measuring its IP remain limited. In most cases, the IP of LAMOST is approximated by a symmetric Gaussian function. However, the actual IP can deviate from a perfectly symmetric profile. Since LAMOST's low-resolution spectra primarily use a mercury–cadmium (Hg–Cd) arc lamp for blue and a neon–argon (Ne–Ar) arc lamp for red as the reference source for wavelength calibration (Z.-R. Bai et al. 2017), we attempt to use these lamp spectra to characterize the IP of LAMOST. Here we used the full width at half-maximum (FWHM), skewness, and kurtosis to illustrate the emission line profiles' asymmetry, as shown in Figure 1, which clearly reveals the profile variations across the detector. Specifically, changes in environmental conditions (e.g., temperature) can induce minor deformations in the mechanical structure and optical mirrors, leading to variations in both the profile shape and its position on the CCD subpixels. Since calibration lamp exposures are not acquired simultaneously with science exposures (Z.-R. Bai et al. 2021), the time interval between them will inevitably introduce uncertainty, as environmental conditions can change over this period. Therefore, using symmetric functions is inappropriate for describing such asymmetric and complexly variable IPs.

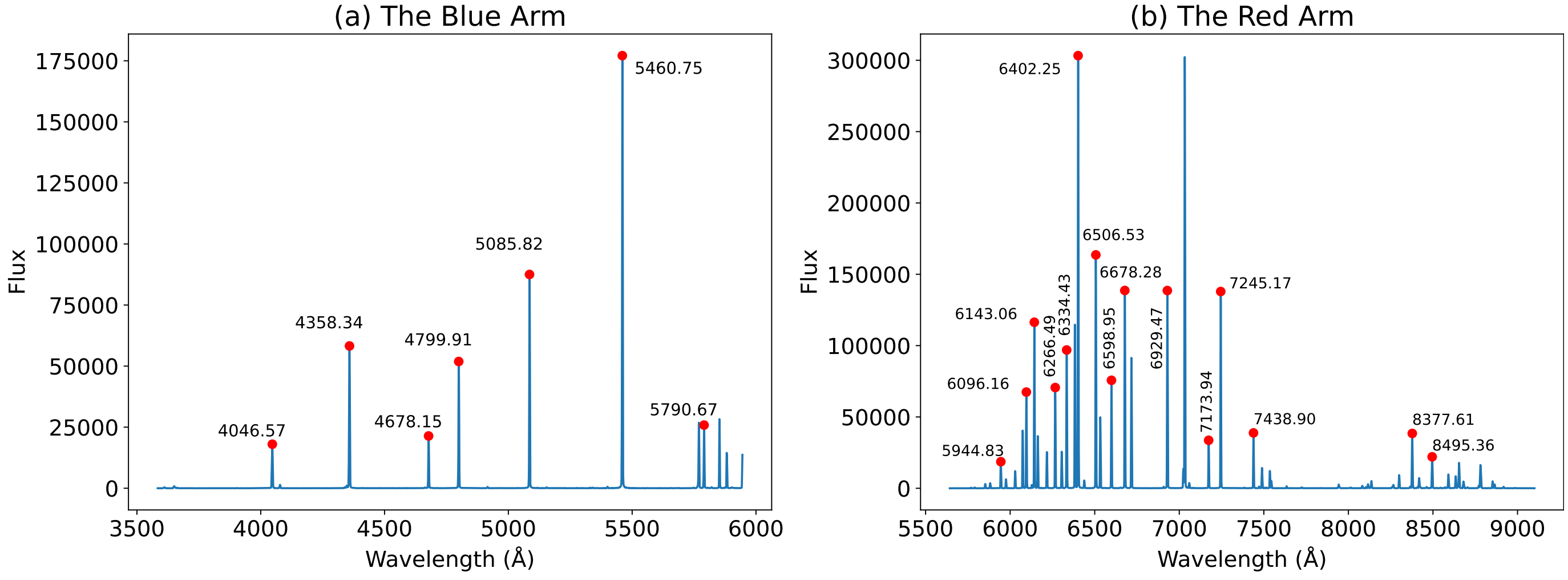

**Figure 2.** Arc lamp spectra used for wavelength calibration of LAMOST low-resolution data. Panel (a) presents an exposure of the Hg–Cd arc lamp used for the blue arm, while panel (b) shows an example of the Ne–Ar arc lamp spectrum obtained for the red arm. The red points mark the peaks of the selected emission lines, with their central wavelengths labeled (in units of Å) adjacent to each point.

Given the critical role of the IP in spectroscopic analysis, accurate characterization of the LAMOST IP will enable the removal of instrumental broadening and distortion, enhance the precision of spectral calibration, and ultimately lay a solid foundation for high-accuracy stellar parameters inference and the identification of rare or peculiar objects, thus fully unlocking the scientific potential of the LAMOST survey.

As mentioned earlier, iodine absorption cells and LFCs can provide dense, relatively uniform, and well-characterized spectral lines, making them more suitable for forward modeling in high-resolution spectrographs to obtain high-precision RV and IP measurements. However, LAMOST is not equipped with an LFC, and the available arc lamps only produce relatively sparse and unevenly distributed emission line profiles. Furthermore, the factors influencing LAMOST's IP are diverse and difficult to formalize. Under these conditions, machine learning methods represent a more practical and effective approach. A neural network approach based on The Payne framework (Y.-S. Ting et al. 2019) can perform nonlinear interpolation in multiple dimensions (e.g., wavelength, fiber number, and time), which is suitable for our case. Thus, we employ it and use the emission lines from the arc lamp to construct an IP model. The rest of this paper is organized as follows. Section 2 describes the data source and how they were processed. In Sections 3 and 4, we present the method we used and analyze the results, respectively. An application in RV measurement is presented in Section 5. Finally, conclusions are summarized in Section 6.

## 2. Data Reduction

We selected all low-resolution calibration lamp spectra obtained by LAMOST since the beginning of its operation, covering a time span from 2011 September to 2024 June. These wavelength-calibrated one-dimensional spectra are outputs of the LAMOST 2D data reduction pipeline (Z.-R. Bai et al. 2017), with an individual spectrum for each fiber sampled as a function of wavelength. Our further processing of these spectra is as follows.

### 2.1. Select Lines

For the selection of emission lines, those with high flux intensities are preferred. Additionally, selected lines should be free from contamination by nearby features and collectively provide broad coverage across the spectral range. The wavelength coverage spans 3700–5900 Å in the blue band and 5700–9000 Å in the red band. Due to the sparsity of lines in the blue band, selection is relatively straightforward. In contrast, the red band exhibits a much higher line density, making the selection process more complex and require visual inspection. Based on these criteria, 7 lines in the blue band and 15 in the red band were chosen, and the final set of selected emission lines is shown in Figure 2, most of which are also used to do wavelength calibration.

### 2.2. Determine a Wavelength Range

A wavelength window was selected around each emission line to fully capture the line profile while avoiding contamination from nearby lines. The widths of different spectral lines vary, and even the same line may exhibit different widths across different fibers (as can be seen from Figure 1). To ensure consistency during data training, the step size of the data points within each line profile must be uniform across the same dataset. Therefore, we evaluated the width of each line and adopted the largest profile range that can encompass all line profiles as well as excluding neighboring lines or noises. Finally, each line profile extends from the central wavelength by $-8.335$ to $+9.25$ Å in the blue arm, while in the red arm, the range is from $-9.163$ to $+9.833$ Å around the central wavelength.

### 2.3. Background Subtraction

Since the selected lines are isolated, the procedure for background subtraction of each line profile is as follows: two local minima were identified on either side of the profile, and a linear fit connecting these two points was used to model the background, which was then subtracted from the flux.

### 2.4. Resample and Normalization

Next, we used spline interpolation to get an emission line profile with a sampling rate about 10 times that of the original lamp spectrum, resulting in 600 data points in the selected wavelength range for each line. For normalization, the peak value of the line profile was defined as equal to 1 to avoid differences in flux intensity. Any negative values following normalization were replaced with 0.

The chosen lines are not always clean as environmental or instrumental effects can produce poor-quality spectra. To ensure the validity of the processed line profiles, the following methods were used throughout this process to eliminate invalid data: a flux threshold was set for each spectrograph's exposure, and fibers with flux below half the median flux were removed to exclude broken or low-flux fibers. We also identified abnormal profile shapes to ensure that the final profiles exhibit nearly single-peaked emission lines, thereby eliminating contamination from cosmic rays and other noise.

### 2.5. Assign Three Features

Apart from the wavelength and flux representing the profile, some other related parameters should also be included. Clearly, the shape of the IP evolves over time and also depends on the wavelength and the position of the fiber. To capture these dependencies, we assigned three features to each line:

1. Observation time—represented as the number of minutes since 00:00:00 on 1858 November 17, which is commonly embedded in the LAMOST spectrum filename.
2. Fiber ID—ranging from 1 to 250, indicating the spatial position of the fiber.
3. Central wavelength—the 22 center wavelengths chosen in Figure 2.

With these features as inputs, the neural network can better learn the nonlinear complex relationship between the IP variations and the physical factors influencing them.

### 2.6. Assemble the Dataset

After processing all the lines, we proceed to assemble the dataset. Since fiber replacements or upgrades may occur during instrument maintenance between observing seasons, which can lead to changes in the IP, data from each observing season should be trained independently. Furthermore, the blue and red arms of each spectrograph function as separate instruments. Therefore, we treat each combination of observing season (typically from September to the following June), spectrograph (Nos. 1 to 16), and the spectral arm (blue or red) as an independent dataset. This means that all line profiles from the same season, spectrograph, and spectral arm were grouped into a single dataset.

Consequently, we got 416 independent datasets. Each dataset contains approximately 0.4 to several million samples, with each emission line contributing up to around 1000 times.

## 3. Method

### 3.1. Network Architecture

The Payne is a machine learning model for the precise and simultaneous determination of numerous stellar labels—such as effective temperature, metallicity, and surface gravity—from observed spectra (Y.-S. Ting et al. 2019). Based on this and the characteristics of our own spectral datasets, we designed a feed-forward neural network to model the mapping between the IPs and their features. The architecture is a multilayer perceptron implemented in PyTorch. The input layer takes the above three features (illustrated in Section 2.5) as inputs, followed by fully connected hidden layers, each with $N$ neurons, and the LeakyReLU activation functions. The output layer is a linear layer producing flux predictions at all spectral wavelength points ($n = 600$ in our case). Formally, the network can be written as (taking a three-layer architecture as an example)

$$y = W_3\,\sigma(W_2\,\sigma(W_1 x + b_1) + b_2) + b_3, \tag{3}$$

where $x$ is the normalized label vector, $\sigma$ denotes the LeakyReLU activation function, and $W_i$ and $b_i$ are the weights and biases of the $i$th layer. The number of layers in the perceptron is to be determined by their performance. In practice, we adjust the depth of the network according to the size of the dataset during training in order to obtain the optimal model.

### 3.2. Training and Inference

Before training, the three inputs were linearly scaled to the range $[-0.5, 0.5]$. During the training of our neural network model, we employed a flux-dependent weighted loss scheme, in which each data point's residual was divided by its corresponding flux value to mitigate fitting inaccuracies in the profile wings, where deviations are prone to being overlooked due to low flux levels. The model optimization was performed using the Rectified Adam algorithm (D. P. Kingma & J. Ba 2015; L. Liu et al. 2019).

For each dataset, samples were split into training and validation sets with an 8:2 ratio. The model was trained in batches over a fixed number of steps, with validation loss calculated every 100 steps to monitor performance and diagnose potential overfitting. The best-performing model was selected based on the lowest validation loss.

After training, the model weights and biases were stored together with the scaling factors used for the inputs. During inference, the scaled inputs were fed into the trained neural network to generate the normalized IP. The forward propagation through the hidden layers with LeakyReLU activations, followed by the linear output layer, produces the flux values across all wavelength points. This inference procedure enables us to predict continuous IP for arbitrary combinations of the inputs.

## 4. Results

We trained each dataset multiple times with different hyperparameter settings such as the number of neurons, the learning rate, and the number of layers in the perceptron. After comparative validation, we decided to adopt a three-layer neural network, with the number of neurons dynamically adjusted according to the size of each dataset. Finally, we got two files for each dataset. One recorded the training and validation loss values computed at regular intervals, and the other saved the weights and biases of the best model as well as the normalization parameters of the three features (i.e., observation time, fiber ID, and central wavelength), helping to get the predicted IP according to the inputs.

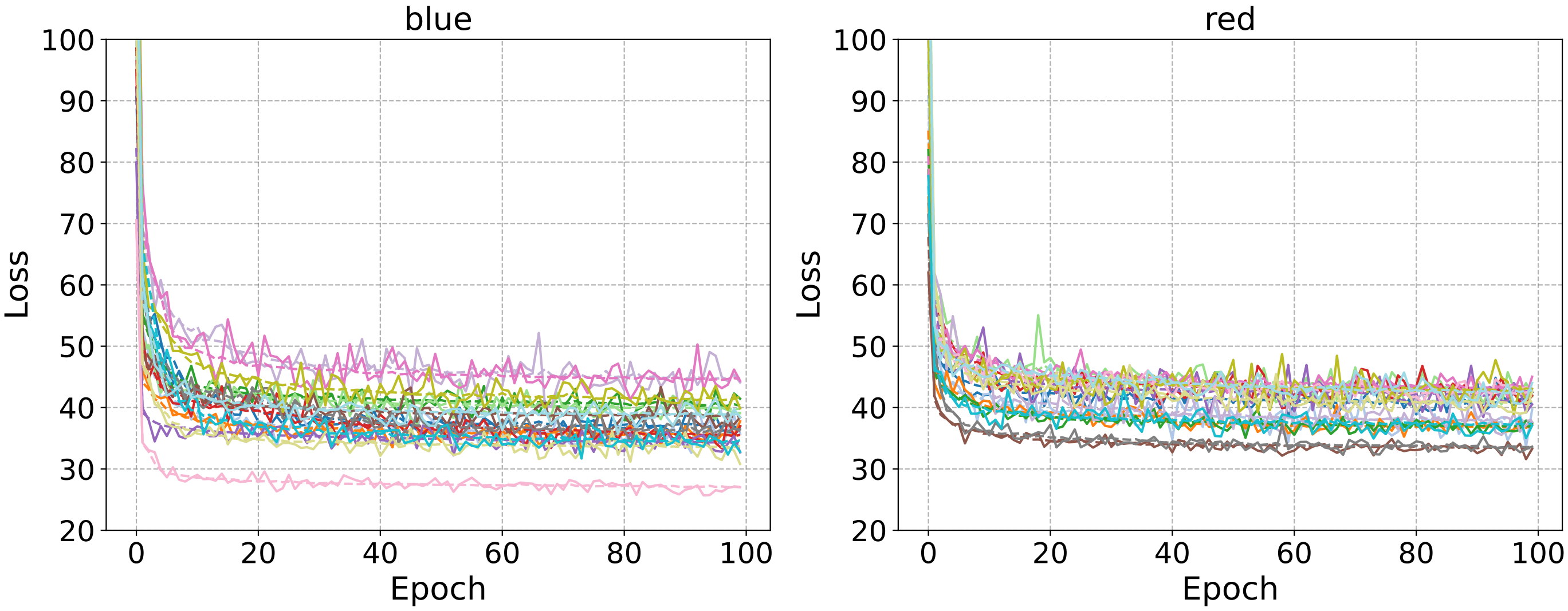

**Figure 3.** One example of the training and validation loss curves in the year of 2024. Colors of the lines indicate the 16 spectrographs of LAMOST, and "blue" and "red" mean the corresponding bands. Solid lines represent the training loss, while dashed lines of the same color correspond to the validation loss.

### 4.1. Residuals

Using the 2024 dataset as an example, the training and validation loss curves are shown in Figure 3. The loss decreases rapidly during the initial epochs and gradually stabilizes, indicating stable convergence of the model. The close agreement between the training and validation losses suggests good generalization performance without significant overfitting.

We evaluated the precision of the neural network in modeling the IP by analyzing the residuals between the observed and predicted profiles, thereby assessing the prediction performance for each sample. The residual of one profile is determined by Equation (4):

$$\mathrm{mean}_{\mathrm{res}} = \frac{1}{n}\sum_{i=1}^{n}(o_i - p_i), \tag{4}$$

$$\sigma_{res} = \sqrt{\frac{1}{n-1}\sum_{i=1}^{n}((o_i - p_i) - \mathrm{mean}_{\mathrm{res}})^2}\,, \tag{5}$$

where $o_i$ denotes the observed value at the $i$ th point, $p_i$ the neural network prediction, and $n$ the length of the IP. Equation (5) shows how the dispersion of one profile's residual is defined. Figure 4 presents the distribution of the residuals and their dispersion for all samples, with the color of the lines indicating different spectral bands. The narrow distribution of the mean residuals and a sharp peak centered near zero indicate that the differences between the predicted and observed IPs are consistently small and exhibit no significant systematic bias. In both the blue and red bands, the model achieves high overall prediction accuracy, with the majority of samples exhibiting mean residuals below 0.005. On the other hand, the average of the residuals dispersion is about 0.006 in the blue arm and 0.007 in the red arm, which implies a stable performance of the model across the entire spectral range—whether for smooth continuum regions or sharp spectral features. Such stability confirms the model's robustness in capturing both global trends and local details of the spectral profiles.

For regions with large residuals, we conducted further analysis and concluded that abrupt changes in the line profile may be responsible. In principle, the IP derived from adjacent fibers should vary smoothly, and our use of the model for interpolation assumes this smoothness. However, when a sudden deviation occurs—where the line profile of a particular fiber differs significantly from its neighbors—the fit of goodness tends to degrade. This results in residuals that are typically 2–3 times larger than those of surrounding fibers, as illustrated in Figure 5. In the 2D fiber map shown in Figure 6, the central fiber (highlighted with a red dashed box) appears to be partially defective compared to its neighbors, supporting this interpretation.

### 4.2. Further Validation

To further assess the performance of our model, we validated its ability to generate line profiles using a spectral line that was not included in the datasets. The [O I] 5577 Å forbidden emission line (hereafter [O I] line) is located between the two training wavelengths 5460.75 and 5790.67 Å. It originates from night-time airglow in the Earth's upper atmosphere and can be reliably recorded by the dedicated sky fibers of LAMOST. This makes it an ideal test case to evaluate the generalization capability of the model by comparing the observed line profiles with the predicted ones. We extracted the [O I] lines from all LAMOST sky-fiber spectra with an exposure time of 30 minutes and carried out the same processing as described in Section 2. For each extracted [O I] line, we generated the corresponding predicted profile using our model and compared it with the observed profile to compute the residuals. Panel (a) in Figure 7 shows one example of the recovered [O I] line profile. For all [O I] line samples, the distributions of their residuals and dispersions are present in panels (b) and (c). It is evident that the prediction performance for the [O I] line is inferior to that of the trained lamp spectral emission lines. Nevertheless, the residuals are basically constrained below 0.03, which represents the optimal result achievable through multiple rounds of testing.

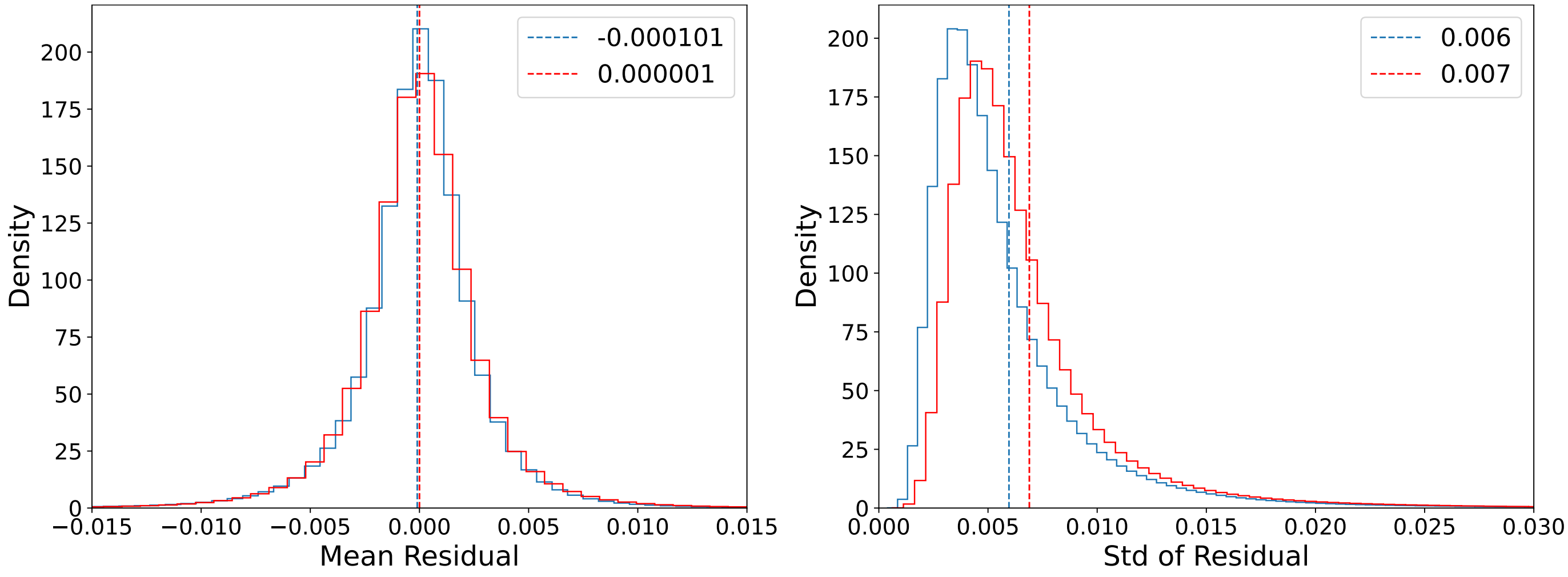

**Figure 4.** The left and right panels show the distributions of $\mathrm{mean}_{\mathrm{res}}$ and $\sigma_{\mathrm{res}}$ for all samples, respectively. The color of the lines indicates the spectral band, and the dashed lines represent the mean value.

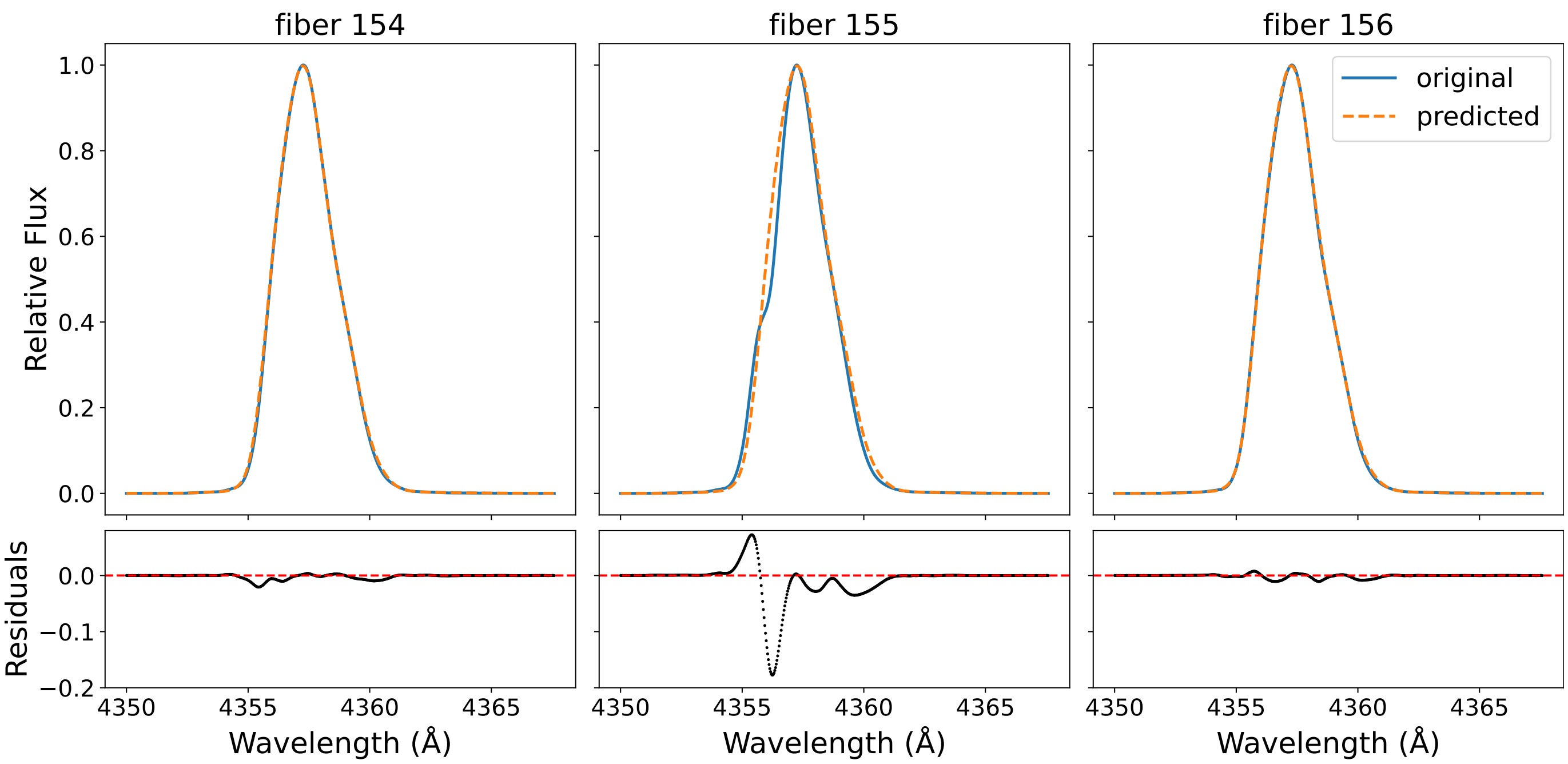

**Figure 5.** Comparison between the original and neural network predicted IP for three adjacent fibers (fiber 154–156). The top panels show the normalized flux profiles, where the blue lines represent the original IP and the orange dashed lines show the predicted IP. The bottom panels show the residuals (original—predicted), with the red dashed lines indicating zero. The middle panel (fiber 155) exhibits significantly larger residuals compared to its neighbors, suggesting a possible abrupt change in the IP shape for that fiber. This deviation likely explains the relatively poor fitting performance for fiber 155.

Finally, we can predict the IP of LAMOST at any given time, for any fiber, and at any wavelength. Of course, the given time should ideally fall within the observation period. In theory, any wavelength can be used. However, some wavelengths do not correspond to real emission lines and are thus meaningless. Others may correspond to actual lines but were not included in the training and therefore may yield unreliable results. The set of wavelengths we trained on is well distributed across the full spectral range and should be sufficient for most practical applications.

## 5. Application

We applied the measured IPs to determine the RV of stars and further evaluate the accuracy of our IP model. The RV was obtained by minimizing the chi-square between the observed and modeled spectra. Specifically, we first convolved a high-resolution template spectrum with the IP, effectively matching it to the resolution of the LAMOST instrument. Both the template and the observed spectra were then normalized using the same continuum-fitting procedure, and cosmic-ray contamination in the observed spectra was removed. The template spectrum was Doppler-shifted over a range of trial velocities, and the velocity corresponding to the minimum chi-square value was taken as the best-fit stellar RV.

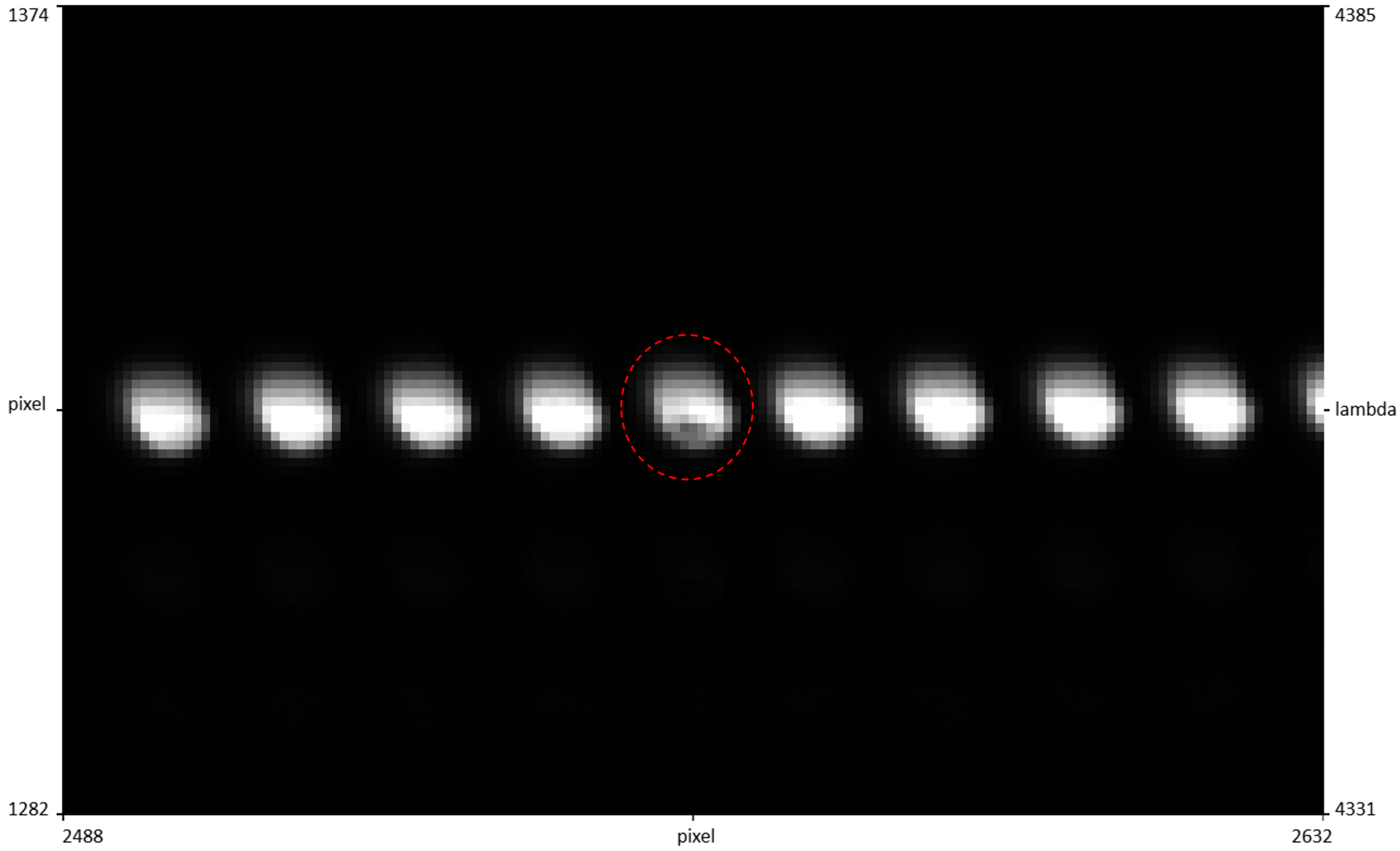

**Figure 6.** Each bright spot is the profile of an individual fiber. The fiber circled in red (fiber 155, corresponding to the middle panel in Figure 5) exhibits an abnormal or incomplete shape compared to its neighboring fibers, suggesting a potential defect of the fiber.

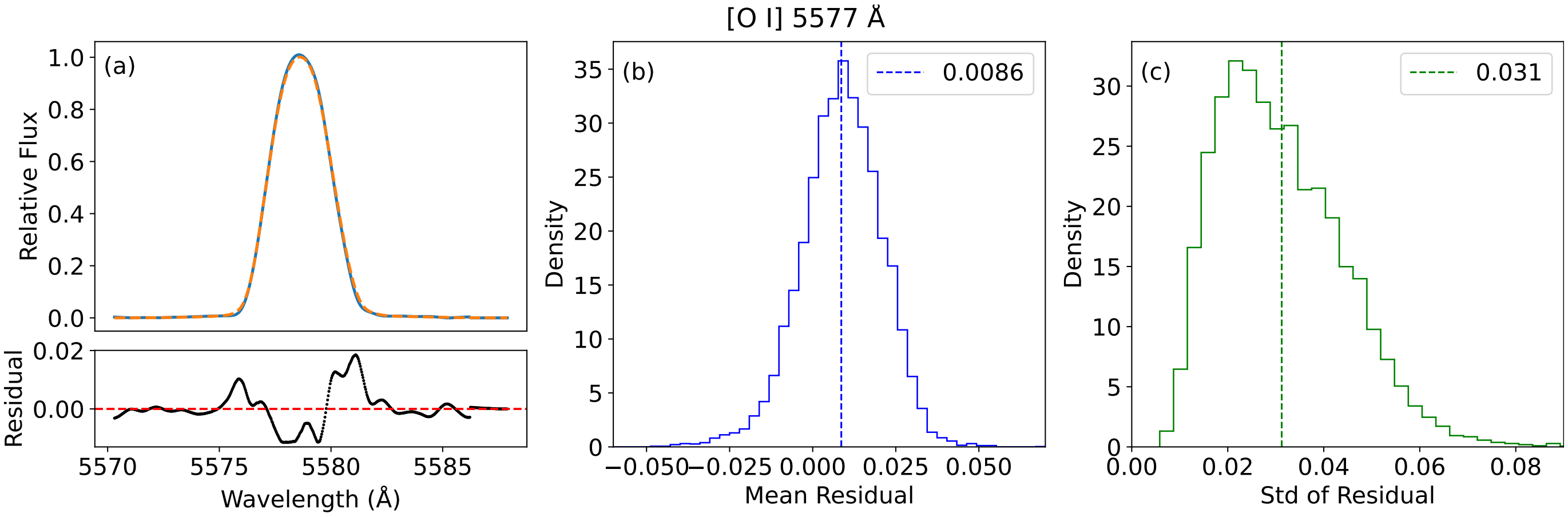

**Figure 7.** Panel (a) shows one example of the recovered [O I] 5577 Å line profile using our IP model. Panels (b) and (c) show the distributions of $\mathrm{mean}_{\mathrm{res}}$ and $\sigma_{\mathrm{res}}$ for all [O I] lines from the year of 2024, respectively. The dashed lines denote the mean value.

## 5.1. Different IP Centers

In the standard LAMOST pipeline, a canonical Gaussian kernel is used whose standard deviation is determined by the instrument resolution and the central wavelength. The width of the Gaussian kernel is typically set to $5\sigma$, where $\sigma$ is calculated from the resolution using the relation $\sigma = \frac{\mathrm{FWHM}}{2.355} = \frac{\lambda}{R \times 2.355}$, where $\lambda$ is the wavelength and $R$ is the resolution power of the instrument.

To assess the accuracy of our IPs, we performed a comparison using both the above Gaussian kernel and our measured IPs for convolution. The Gaussian kernel, being symmetric and parameterized by the instrumental resolution, does not account for asymmetries in the actual line profiles. In contrast, the measured IPs, derived from the arc lamp exposures, often exhibit asymmetry. Therefore, determining the central position—or velocity zero-point—of these asymmetric IPs becomes nontrivial as it can significantly affect the derived RV (D. Milaković & P. Jethwa 2024). To investigate this, we adopted six different definitions for the IP center:

1. *Max.* The position of maximum intensity of the IP.
2. *Centroid.* The flux-weighted centroid of the IP.
3. *Gauss.* The center of a Gaussian profile fit to the IP.
4. *Sérsic.* The symmetric center obtained by fitting the IP with a Sérsic profile (Z.-R. Bai et al. 2017), which is in the form

$$f(\lambda) = \alpha e^{-\frac{|\lambda - \lambda_i|^{\delta}}{\delta_{\gamma}^{\delta}}} + a\lambda + b, \tag{6}$$

where $f(\lambda)$ is the relative flux, $\alpha$ and $\delta$ are parameters of the Sérsic function, and $a\lambda + b$ is the fitted background continuum.

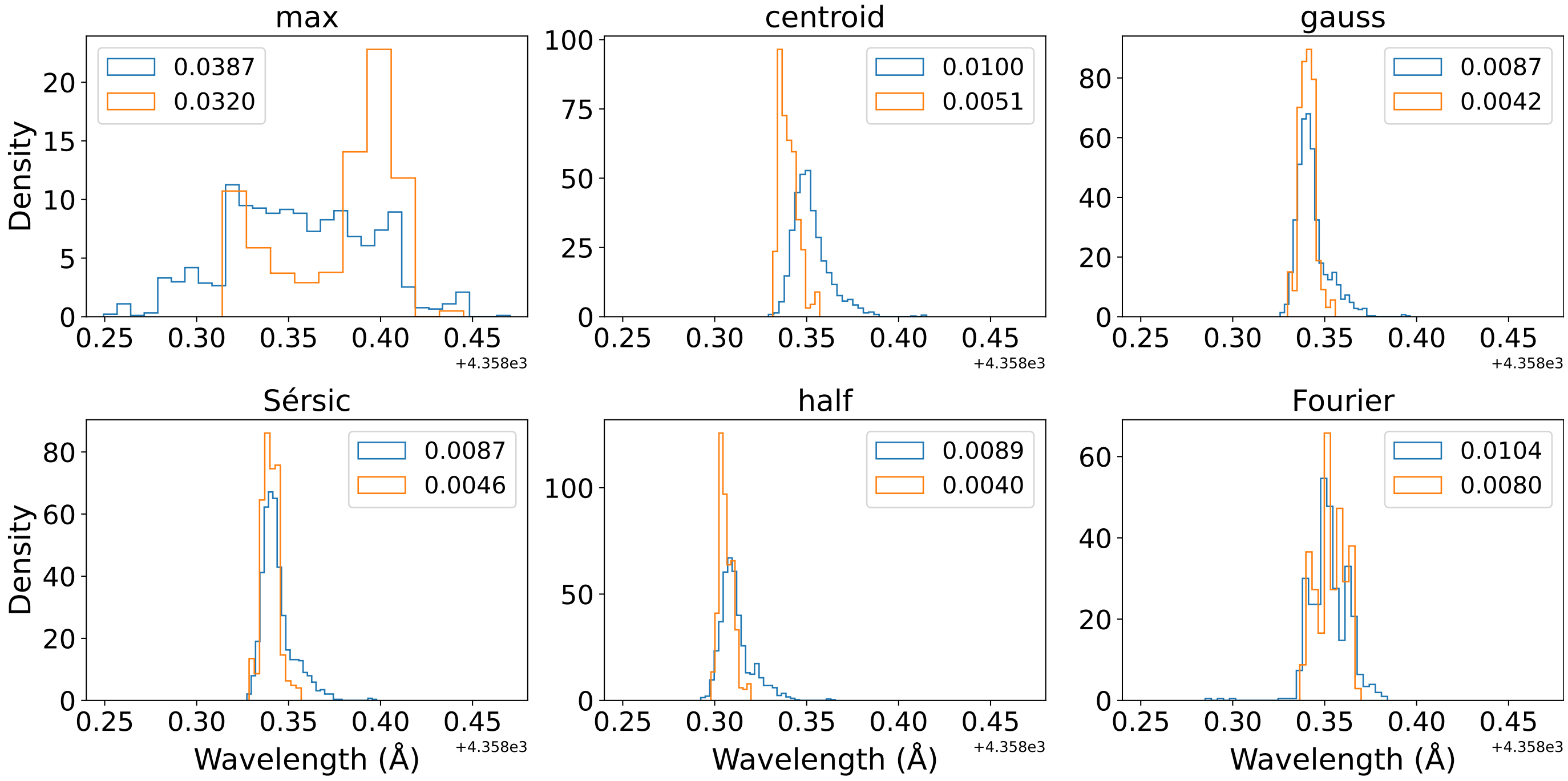

**Figure 8.** Distributions of the centers of both the original (blue) and neural network predicted (orange) line profiles determined by the above six definitions, as indicated by the titles of each panel. The profiles used here are the emission lines with the central wavelength of 4358.34 Å imaged by the first fiber of Spectrograph No. 1 in 2017. The numerical values in the figure represent the standard deviations of the distributions.

5. *Half.* The point at which the integrated flux on both sides of the IP is equal (J. Anderson & I. R. King 2000; D. Milaković & P. Jethwa 2024).
6. *Fourier.* A weighted center based on a truncated Fourier decomposition of the IP, where each component's center is weighted by its amplitude.

We calculated the centers of both the original and neural network predicted line profiles determined by the above six definitions, as shown in Figure 8. It is evident that using the peak position as the center is the least stable approach, since even slight variations in the profile shape can significantly shift the peak, resulting in a larger dispersion. Therefore, we consider that the RV measurements based on the peak position may be less reliable than those derived from other methods. Furthermore, the figure demonstrates that the centers of the neural network predicted profiles are more stable, exhibiting smaller dispersion compared to those of the original profiles. This enhanced stability implies that neural network predictions can serve as a robust basis for RV determination, reducing the impact of noise and subtle distortions in the observed spectra.

### 5.2. RV Measurement of a Star

As a test case, we selected a star with R.A. = $93\overset{\circ}{.}9960641$ and decl. = $23\overset{\circ}{.}2585934$. The Gaia parameters of this target (L. Lindegren et al. 2021; Gaia Collaboration et al. 2023) are listed in Table 1. It is classified as an F7-type star with an effective temperature of approximately 6239 K. The observed spectra from the LAMOST low-resolution spectrograph span from 2016 November to 2023 January, with a total of 24 observation sessions and 230 exposures. To ensure reliable RV measurements, we selected only those exposures with signal-to-noise ratio (S/N) $\geqslant$20, resulting in 179 valid spectra with an average S/N of about 29. These spectra are distributed as follows: 83 exposures in 2016 (November and December), 70 in 2017 (January, February, November, and December), 5 in December 2021, 16 in 2022 (February, March, and October), and 5 in January 2023.

**Table 1**
Main Parameters of the Target

| Parameter | Value |
|---|---|
| source id | 3425173819415139328 |
| r.a. | 93.99612278582052 |
| decl. | 23.258594838470284 |
| parallax | 1.298950651161706 |
| parallax error | 0.016847283 |
| RUWE | 1.1137948 |
| RV | 31.926884 |
| RV error | 2.633911 |
| pmra | −0.8408199097098139 |
| pmra error | 0.017269075 |
| pmdec | −1.064246912284117 |
| pmdec error | 0.013009709 |

For each spectrum, we computed the RV using each of the six IP center definitions and the commonly used Gaussian kernel. When using our measured IPs for convolution, the IP was selected according to the observation time and fiber ID. For each RV measurement, we extracted a spectral segment centered on the central wavelength of the corresponding IP, extending ±150 Å on either side.

Figure 9 displays the results for a spectral segment centered at 4358.34 Å. We can see from the figure that the improvement gained from using our measured IPs instead of a Gaussian function is marginal and barely visible. This is expected: a well-trained neural network is not supposed to generate predicted IPs that deviate drastically from the original ones. Consequently, the RVs computed using our IPs show only minor differences compared to those derived from the Gaussian approximation.

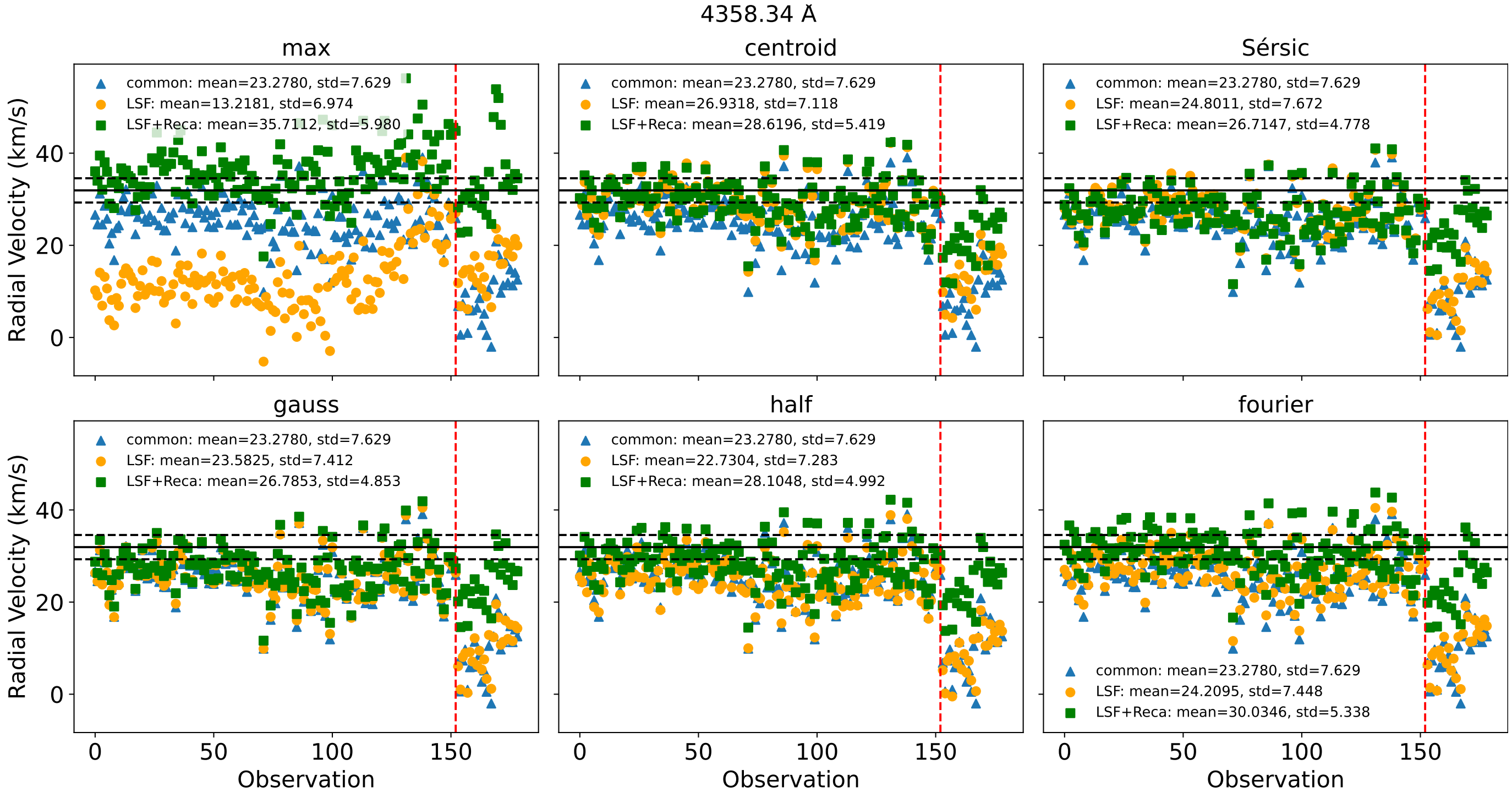

**Figure 9.** RV results for a spectral segment centered at 4358.34 Å. Each point represents the RV computed for one exposure, plotted in chronological order. The blue triangles indicate RVs computed using a commonly adopted Gaussian profile, the orange circles represent results derived using our measured IPs, and the green squares show results using our IPs after recalibrating the wavelength solution. The titles clarify the central wavelength of the used IPs and how they were centered. The numbers in the legend indicate the mean and standard deviation of the velocity measurements. The black solid and dashed lines denote the Gaia RV and its associated error from Table 1, respectively. Since these spectra were taken at different times, with the maximum time gap being approximately 4 yr, the red dashed lines mark this boundary.

One noticeable feature in the figure is a clear discontinuity at the position marked by a vertical dashed line, showing a sudden jump of over 10 $\mathrm{km\,s^{-1}}$. The Renormalized Unit Weight Error (RUWE) of 1.1137948 in Table 1 indicates that the star's positional measurements are consistent with a single-star model. We also cross-matched the source with the Gaia DR3 Non-Single Star catalog (Gaia Collaboration et al. 2023) and found no corresponding entry. Additionally, the star has 15 observations, and its RV uncertainty is only $\sim$2.6 $\mathrm{km\,s^{-1}}$, suggesting that the likelihood of unseen companions significantly affecting its RV is low. Therefore, the observed jump is unlikely to be caused by the star's intrinsic motion. Upon investigation, we found that the closest exposures on either side of the dashed line were taken in 2017 December and 2021 December, respectively. This indicates that the data obtained before or in 2017 December differ from those collected in or after 2021 December by approximately 10 $\mathrm{km\,s^{-1}}$. Since the convolution kernel plays a crucial role in RV measurement, we conducted a further analysis of the IP and found that the emission line profiles used for wavelength calibration and resolution-degrading convolution differ substantially between the two epochs, as shown in Figure 10. Compared to the profiles from 2017 and earlier, those from 2021 onward have not only become wider but also undergone significant changes in shape.

### 5.3. RV Measurement after Wavelength Recalibration

Since such changes in these profiles can also affect wavelength calibration, we used our IP model to perform a recalibration of the wavelength solution. For each exposure, the temporally nearest arc lamp exposure was identified and used for wavelength calibration. The central pixel positions of the arc lamp emission lines were determined and then associated with their corresponding known rest-frame wavelengths. The resulting set of pixel–wavelength pairs was used to fit a fifth-order polynomial or a Legendre polynomial, thus establishing a continuous wavelength solution for the spectra. Our recalibration differs from the default LAMOST pipeline in one key way: while the pipeline determines the center of each line by fitting a Sérsic profile and finding its symmetric center, we used the six methods for determining the velocity zero-points of IPs discussed earlier to locate the pixel centers. In other words, we applied a consistent method to determine both the velocity zero-points of the IPs and the pixel positions of the arc lamp emission lines used in wavelength calibration.

After recalibrating the wavelength solution, we recomputed the RVs using the measured IPs. The green squares in Figure 9 show the updated results. The previously observed discontinuity largely disappears, and the RV scatter decreases by approximately 1–3 $\mathrm{km\,s^{-1}}$. This improvement likely stems from better alignment between the stellar spectrum and the template under a consistent IP, reducing systematic differences caused by temporal variations in arc lamp profiles. Compared with the RV from Gaia, the differences are minor: the mean of our measurements deviates from the Gaia value by no more than 2 $\mathrm{km\,s^{-1}}$, indicating good consistency. According to Table 4 in Z.-R. Bai et al. (2021), the standard deviation of RV between two exposures on different days for this stellar type and signal-to-noise ratio is about 6.19 $\mathrm{km\,s^{-1}}$. Obviously, our measurements are below this value. Since the results obtained using various methods are consistent, we consider this approach to be reliable in reducing the RV dispersion.

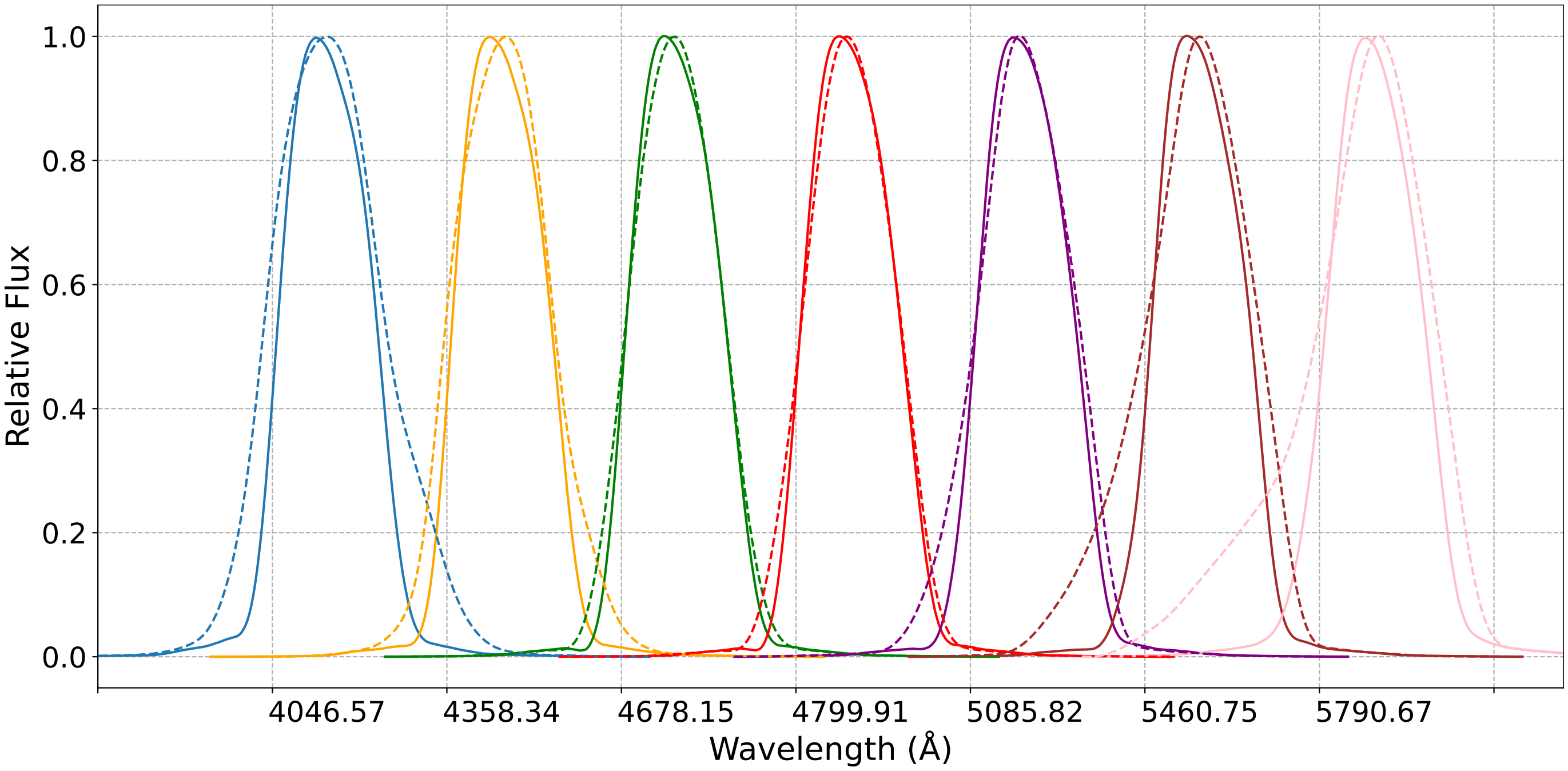

**Figure 10.** The predicted IPs on two dates. From left to right, the IPs correspond to central wavelengths of 4046.57, 4358.34, 4678.15, 4799.91, 5085.82, 5460.75, and 5790.67 Å. Solid lines indicate profiles from an exposure on the left side of the red dashed line in Figure 9, while dashed lines correspond to ones on the right side.

## 6. Conclusion

LAMOST's arc lamp spectra provide a clear demonstration of how an ideal light source is broadened and distorted by the instrument. In this work, we utilized all available low-resolution arc lamp spectra collected since the telescope's commissioning and trained a neural network model based on The Payne to measure the LAMOST's IP. The trained model enables the reconstruction of the IP for any fiber, at any wavelength and time. For trained lines, the accuracy is constrained below 0.5% peak intensity. For lines not included in the training data, the model can recover the [O I] 5577 Å line with residual below 3% peak intensity. The resulting IPs have been encapsulated into a Python package, available for installation and utilization through download from GitHub[8] or the Zenodo archive doi:10.5281/zenodo.18667269 (Q. Liu et al. 2026).

To validate the IP, we used it as a convolution kernel to degrade template spectra and recalibrated the wavelengths of stellar spectra, subsequently computing RVs. During this process, six methods were employed to determine the center of the IP. The preliminary results show that, except for the method using the IP peak, all approaches reduce RV dispersion by 1–3 $\mathrm{km\,s^{-1}}$, demonstrating the reliability of the measured IPs. This improvement has the potential to facilitate the detection of long-period binary systems through RV variations.

Building on these results, future work will focus on applying the reconstructed IPs to large-scale analyses of LAMOST low-resolution spectra, where accurate IP characterization is essential for precise measurements of stellar parameters and chemical abundances. Large-scale application will also enable a more comprehensive validation of the model performance. In addition, the methodology presented here can be naturally extended to medium-resolution spectroscopy. We also plan to adapt the neural-network-based framework to model the IP of LAMOST medium-resolution spectra, thereby providing a unified and data-driven approach to IP reconstruction across different spectral resolutions.

## Acknowledgments

This work is supported by the National Natural Science Foundation of China (grant Nos. 12273056 and 12090041) and the National Key R&D Program of China (grant No. 2022YFA1603002). H.-L.Y. and Z.-R.B. are supported by the National Key R&D Program of China (grant No. 2023YFA1607901). H.-L.Y. also acknowledges support from the Youth Innovation Promotion Association of the CAS (grant No. 20200060), the National Natural Science Foundation of China (grant No. 11873066), and the Strategic Priority Program of the Chinese Academy of Sciences (grant No. XDB1160302). M.Z. is supported by the National Natural Science for Youth Foundation of China (grant No. 12503091).

This work has made use of data products from LAMOST, which is operated and managed by the National Astronomical Observatories, Chinese Academy of Sciences (http://www.lamost.org/public/?locale=en). Funding for the construction and operation of LAMOST has been provided by the National Development and Reform Commission of China.

## ORCID iDs

Qian Liu (刘倩) https://orcid.org/0009-0009-0971-5081
Zhongrui Bai (白仲瑞) https://orcid.org/0000-0003-3884-5693
Ming Zhou (周明) https://orcid.org/0000-0003-1487-0093
Mingkuan Yang (杨明宽) https://orcid.org/0009-0000-6595-2537
Xiaozhen Yang (杨肖振) https://orcid.org/0009-0006-7506-1299
Ziyue Jiang (蒋子悦) https://orcid.org/0009-0009-9065-1846

[8] https://github.com/catty215/Construct_LAMOST_IP.git

Hailong Yuan (袁海龙) https://orcid.org/0000-0002-4554-5579
Ganyu Li (李甘雨) https://orcid.org/0009-0009-0171-1553
Yuji He (何玉吉) https://orcid.org/0009-0008-0146-118X
Mengxin Wang (汪梦欣) https://orcid.org/0009-0009-6931-2276
Yiqiao Dong (董义乔) https://orcid.org/0000-0002-6312-4444
Haotong Zhang (张昊彤) https://orcid.org/0000-0002-6617-5300